# Talent hat, cross-border mobility, and career development in China

Yurui Huang[1], Xuesen Cheng[2], Chaolin Tian[1], Xunyi Jiang[1], Langtian Ma[1], Yifang Ma[1*]

[1]Department of Statistics and Data Science, Southern University of Science and Technology, Shenzhen 518055, Guangdong, China

[2]Center for Higher Education Research, Southern University of Science and Technology, Shenzhen 518055, Guangdong, China

*Email: mayf@sustech.edu.cn

**Abstract** This study aims to investigate the influence of cross-border recruitment program in China, which confers scientists with a "talent hat" including a startup package comprising significant bonuses, pay, and funding, on their future performance and career development. By curating a unique dataset from China's 10-year talent recruitment program, we employed multiple matching designs to quantify the effects of the cross-border recruitment with "talent hat" on early career STEM scholars. Our findings indicate that the cross-border talents perform better than their comparable contenders who move without talent hats and those who do not move, given equivalent scientific performance before relocation. Moreover, we observed that scholars in experimental fields derive greater benefits from the talent program than those in non-experimental fields. Finally, we investigated how the changes in scientific environment of scientists affect their future performance. We found that talents who reassembled their collaboration network with new collaborators in new institutions after job replacement experienced significant improvements in their academic performance. However, shifting research directions entails risks, which results in a subsequent decrease of future productivity and citation impact following the relocation. This study has significant implications for young scientists, research institutions, and governments concerning cultivating cross-border talents.

# 1. Introduction

Scientific career movement is fundamental for science advances, which not only accelerates the circulation rate of knowledge across institutions and national borders but also enriches the scientific curriculum for researchers ([Trippl 2013](), [Deville, Wang et al. 2014](), [Petersen 2018](), [Verginer and Riccaboni 2021]()). Studies showed that nations with scientific openness to international mobility and collaboration are linked to stronger research ([Wagner and Leydesdorff 2005](), [Sugimoto, Robinson-García et al. 2017](), [Wagner and Jonkers 2017]()). In recent decades, many countries have taken actions to impel cross-border talent recruitment. For instance, the Young Thousand Talents program in China ([Jia 2018](), [Yang and Marini 2019]()) in cultivating and recruiting potential future rising star scientists overseas, and it has attracted more than three thousand young talents over the world. The elected scholars (with "talent hat") will get a big bonus and startup funding from the government and the extra bonus, funding, PhD student quota from the employer university. Besides China, many countries have developed specific Visas for attracting foreign talents, for instance, the "EB-1" in the US and the "Researcher Visa" in Germany.

However, scientists often experience dilemmas in career movements, especially cross-border mobility([Jia 2018](), [Petersen 2018](), [Xu, Braun Střelcová et al. 2022]()): on one hand, career mobility will bring new research opportunities in new environments, collaborators, and maybe a startup funding for rebuilding the lab, which scientists could benefit a lot; on the other hand, mobility may cause short-term research discontinuity and the losing parts of resources from the previous institution, collaborators, and funding agencies. In China, the organizational structure often involves "big team", i.e., one super PI leads a team of multiple investigators. This creates a unique dynamic for younger researchers, particularly recent PhD graduates or postdocs, who must transition to independence and relocate to establish their own scientific careers. Consequently, they face the intricate challenge of balancing family, research, and relocation, which remains a formidable task([Malmgren, Ottino et al. 2010](), [Azoulay, Ganguli et al. 2017](), [Lienard, Achakulvisut et al. 2018](), [Ma, Mukherjee et al. 2020]()).

This prompts us to ask two questions: How do scientists benefit from international talent recruitment, and how to maintain a stable and continuable scientific career after cross-border mobility? In recent decades, there are increasing discussions about talents, impact, career, and mobility, including the statistical modeling of mobility patterns, career development, policy implications ([Deville, Wang et al. 2014](), [Clauset, Arbesman et al. 2015](), [Way, Morgan et al. 2017](), [Petersen 2018](), [Zhang, Deng et al. 2019](), [Cao, Baas et al. 2020](), [Netz, Hampel et al. 2020](), [Zweig, Siqin et al. 2020](), [Cao and Simon 2021](), [Vásárhelyi, Zakhlebin et al. 2021]()), etc. Shi et al. found out that the Young Thousand Talents Program in China (talent with cross-border movement) are successful in attracting high academic caliber but not the top ones compared to those who get the talent hats without returning to China ([Shi, Liu et al. 2023]()), Zhao and Cao et al. found that the returnees with cross-border mobility did not show advantages on scientific productivity ([Zhao, Wei et al. 2023]()) but were more likely to publish higher impact works and more internationally active than domestic counterparts (Cao, Baas et al. 2020), Zweig et al. found that domestic policies are crucial to attracting abroad scholars (Zweig, Siqin et al. 2020). Further, the career development of moved scientists is controversial. Zweig et al. found that scholar with oversea PhDs benefits more in terms of people's perceptions and technology transfer (Zweig, Changgui et al. 2004) and produce more disruptive works (Zhao, Li et al. 2019), while Tang et al. found that Chinese international returnees are in a pessimistic status in career promotion (Li and Tang 2019).

To quantitatively clarify the effects of cross-border movement and talent hat on future scientific performance, we manually collected 10ys of the scholars who were elected to the Young Thousand Talents program in China and manually matched their publication, citation, and collaboration records via a large-scale scientific corpus. The challenge to conduct this research is how to differentiate the indigenous factors which influence the future status of researchers other than mobility, for example, prior scientific impact will predict future performance for scientists. It has been a difficult problem for a long time to obtain persuasive results and unbiased estimation in this comparison because treated group is incomparable with control group in the aspects of group size and prior academic attributes. In recent decades, the matching designs are used in observational data to reduce the confounding influence in science of science and bibliometrics, including topics in gender disparities, citation prediction, mobility, prizewinning, team performance, field growth, collaborations, etc. (Azoulay, Ganguli et al. 2017, Reschke, Azoulay et al. 2018, Jin, Ma et al. 2021, Huang, Tian et al. 2023, Shi, Liu et al. 2023, Zhu, Jin et al. 2023) and the development of network tools to studies scientific careers and science (Wang, Song et al. 2013, Zeng, Shen et al. 2017, Fortunato, Bergstrom et al. 2018, Way, Morgan et al. 2019, Yang, Chawla et al. 2019). In general, the matching techniques are used to find observational "twins" before the treatment and estimate the treatment effects by comparing the posterior difference between them.

In this work, we investigated about 2.6 million scholars with more than 10 publication records and their 65.1 million papers from 2000 to 2021 using OpenAlex (Priem, Piwowar et al. 2022) and manually curated 1,563 Young Thousand Talents and their publication records, which enable us to track the future scientific performance (Petersen, Riccaboni et al. 2012, Sinatra, Wang et al. 2016) for each scholar. We constructed a comparison on three groups of scholars: (1) $G_w$: cross-border scientists with talent hat, i.e., scientists who moved to China and were elected to the Young Thousand Talent program who got large pay, funding, and other resources. (2) $G_1$: cross-border scientists without talent hat, i.e., scientists who moved to China who were not elected to the talent program (See supplementary Fig. S3). (3) $G_2$: non-cross-border scientists, i.e., scientists who have not had cross-border movements (See supplementary Fig. S4). We leveraged the benefits of multiple matching techniques by using the coarsened exact matching (CEM)-like procedure to identify identical scientists on a series of observational variables and the synthetic control method (SCM) (Abadie and Gardeazabal 2003, Abadie,

Diamond et al. 2010, Abadie, Diamond et al. 2015) to further improve the quality of matching in evolving trends.

We found the advantages of cross-border talents in future scientific career success by disciplines and years. However, not everyone benefits the same from the cross-border talent program. Indeed, some of them thrived fast and grew into top scientists among their peers, yet some of them didn't. We further explored the potential environmental changes of the talents and the matched contenders and found that scientists who reassemble their collaboration network with new collaborators in new institutions after the cross-border movement will significantly improve their academic performance while changing research directions after movement may have risk in citation gain, which raise strong policy implications on cultivating cross-border talents.

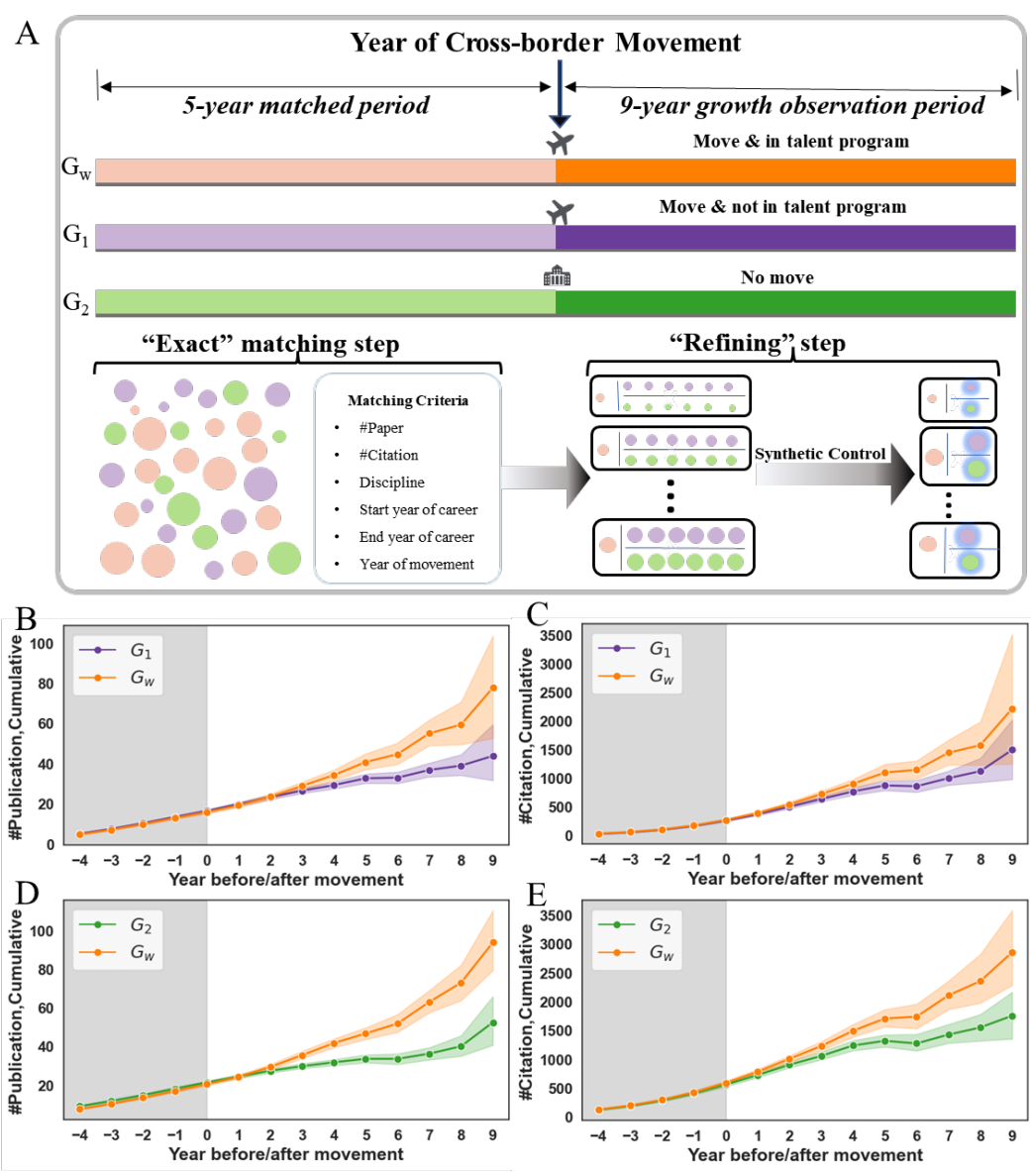

**Figure 1. Two-step procedure for matching groups of contenders. A.** Illustration of the two-step method for matching comparable contender groups who moved without talent hat (G$_1$) and the contender group who did not conduct cross-border movements (G$_2$) before their movements. For each talent, the (coarsened-) "exact" matching step will match contenders with the same discipline, close research career starting year, and the similar total number of publications and citations, then the "refining" step will further match the yearly number of publications and citations to improve the matching precision. **B-E**, show the cumulative number of publications and citations by year for the talents (G$_w$) and the contender groups (G$_1$ & G$_2$).

## 2. Results

**2.1. The talent hat and research designs.**

Since 2011, China's central government had announced the Young Thousand Talent program to attract younger rising star scientists oversea to China, the elected scholar (conferred with a talent hat/title) will be given strong bonus and startup fundings from the government, and the extra bonus, fundings, PhD student quota from the employer university, which is a significantly higher compensation package than scholar not being elected. We manually collected the 1563 talents who enrolled into the program and come back to China successfully in seven cohorts, 2011-2013, 2015-2018 (see section 1 in SI for details).

To quantify the cross-border mobility effects and the talent hat effects, we compared the talents with scientists who moved to China without talent hat (not elected by the talent program, denoted as G$_1$) and scientists who did not conduct cross-border movements (G$_2$). Specifically, as shown in Fig. 1A, we conducted a two-step matching procedure to identify indistinguishable control groups from the millions of scientists in the database. At first, the "exact" step, we conducted a 1:N match, i.e., for each cross-border talent scientist, we matched at most 300 unmoved scientists and 200 moved (to China) scientists who have the same discipline, close research career starting year, and comparable total number of publications and citations with the talent scientist. Second, the "refining" step, we further matched the identical scientists by considering the yearly publications and yearly citations to make sure they display the same career development curve prior to the talent recruitment year (Supplementary Figure S5, Table S2, Table S3). In this step, we leveraged the established matching techniques, the synthetic control method (SCM) in the main results. The SCM will assign weights to each of the talent's contenders from step one to better match the talent scientist's yearly publication and yearly citation curves and can account for the effects of confounders changing with time (The Method section reports SCM details and corresponding equation). Eventually, we got 555 and 588 pairs on publication match and citation match respectively for the talents and the moved scientists without talent hats, 1208 and 1433 matched pairs of scientists on publication match and citation match respectively for the talents and the unmoved. As expected, the matching pairs are fewer a bit for the talents and the moved scientist without talent hats, indicating that the talents elected by the program are top-tier scientists who are not easy to match by scientists missed the program. We also used the coarsened exact matching and the dynamic optimal matching method respectively to validate our results (details of the procedures are in Section 2 of SI).

## 2.2. Moved young scientists with talent hats are successful in career initials.

In Fig. 1B and 1C, we showed the yearly number of publications and yearly number of citations for the talents and the contenders who moved to China without the talent hat, respectively. Indeed, according to the SCM method, they showed the same curves before the movement as shown in the grey areas. After the movement year, the talents' numbers of publications and citations exceed the group without talent hats. The same results are confirmed from both CEM and DOM matching methods (see Table. S3 in SI). In Fig. 1D and 1E, we showed the comparison of the talents and the scientist without cross-border movement, the talents also present higher number of publications and citation impacts in the future after the mobility.

Quantitatively, we performed the difference-in-difference (DID) regression models to further explore the effects of mobility and talent hat, with the control of the fixed effects from the individual variations and time cohorts. As shown in Table 1, the talent scientists published approximately one more paper on average per year after movement and received about 18 percent ($\approx e^{0.1662} - 1$) more citations per year than scientists who move without talent hat. We also found that the talents also performed better than the scientists who did not move cross border, with on average 2 more papers per year and 34 percent ($\approx e^{0.2929} - 1$) more citations.

**Table 1.** DID regression results comparing the talent group ($G_w$), group of scientists who moved without talent hats ($G_1$), and group of scientists who did not move ($G_2$). The regression coefficients are presented with significance levels and the standard errors in parentheses, fixing effects for individual and cohort years.

|  | $G_w$ vs. $G_1$ | | $G_w$ vs. $G_2$ | |
|---|---|---|---|---|
| Scientific performance | Publications | Citations | Publications | Citations |
| Talent hat×Movement | 0.9028*** (0.2033) | 0.1662*** (0.0478) | 2.1166*** (0.1428) | 0.2929*** (0.0277) |
| Individual | Yes | Yes | Yes | Yes |
| Time | Yes | Yes | Yes | Yes |
| #Pairs matched | 555 | 588 | 1208 | 1433 |
| N | 10602 | 11804 | 22725 | 28275 |
| $R^2$ | 0.5036 | 0.8229 | 0.5720 | 0.8586 |

*p<0.05, **p<0.01, ***p<0001

**The talent hat effects are enlarging with career years.** Previous study shows that incentives in early career stage will improve the scientist's future performance (Bol, de

Vaan et al. 2018, Zhu, Jin et al. 2023). Analogously, based on the DID models, we evaluated the yearly differences between the talents and the contenders, as shown in Fig. 2, the talent scientists began to thrive and made more achievements in scientific research compared with both contenders who came back to China without talent hats and contenders who did not conduct cross-border movements. In general, in the initial years after movements, the scientists might be occupied with settling down and reconstructing their research team like recruiting students and research assistants or purchasing equipment, their scientific performance remained similar, the scientific impact difference will increase with time in the first 5-6 years after conferring the talent hat. On average, scientists in $G_w$ published approximately 3 more papers than scientists in $G_1$ in the sixth year after coming to China.

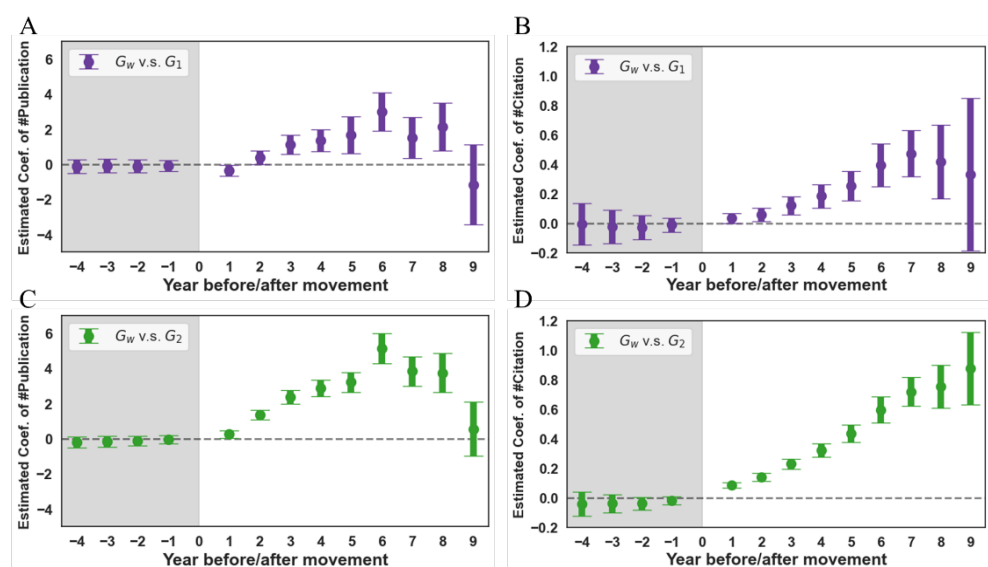

**Figure 2. Estimated difference of publications and citations for the talents and the matched counterparts. A-B**, the annual coefficients for the talents ($G_w$) versus the contenders who moved to China without talent hat ($G_1$). **C-D**, the same estimation for $G_w$ versus the contenders without movements ($G_2$). Dots represent the coefficients, and the error bars denote the 95% confidence intervals.

**Talents from more experimental disciplines gain more benefits.** For younger scientists, start-up fundings are important to build their labs, including recruitment of PhD students, purchase of experimental equipment and materials, technical and environment maintenance, etc. The elected talents will be given sufficient startup fundings from the government and the extra fundings, PhD student quota from the employer university. Indeed, in Fig. 3A and 3B, when we compared the talents with their contenders by disciplines, we found that, in general, the talents benefit more in experimental fields, which need substantial funding support, such as biology, chemistry, computer science, and materials, while the gap is smaller in theoretical fields like mathematics (See Fig. 3A, 3B and Supplementary Fig. S6).

**Talent halo is diluting with time.** When we checked the talent benefits by different year cohorts, in Fig. 3C and 3D, we saw a decrease in the talent benefits compared to the two groups of contenders. Two potential reasons account for this result. On the one hand, in recent years, with the number of returnees increasing, many Chinese universities also launched local parallel talent programs and offered enticing compensation packages to attract younger scholars studying abroad who do not enroll in the talent program. Therefore, the privileges of talents are diluting with time. On the other hand, the benefit gaps between the talents and the contenders are increasing with time based on Fig. 2, so we do not have enough observational years for scientists in recent cohorts, for instance, for scientists in the 2018 cohorts, we only witnessed less than 3 years (as of 2020 for our database) after they moved to China (Supplementary Fig. S7).

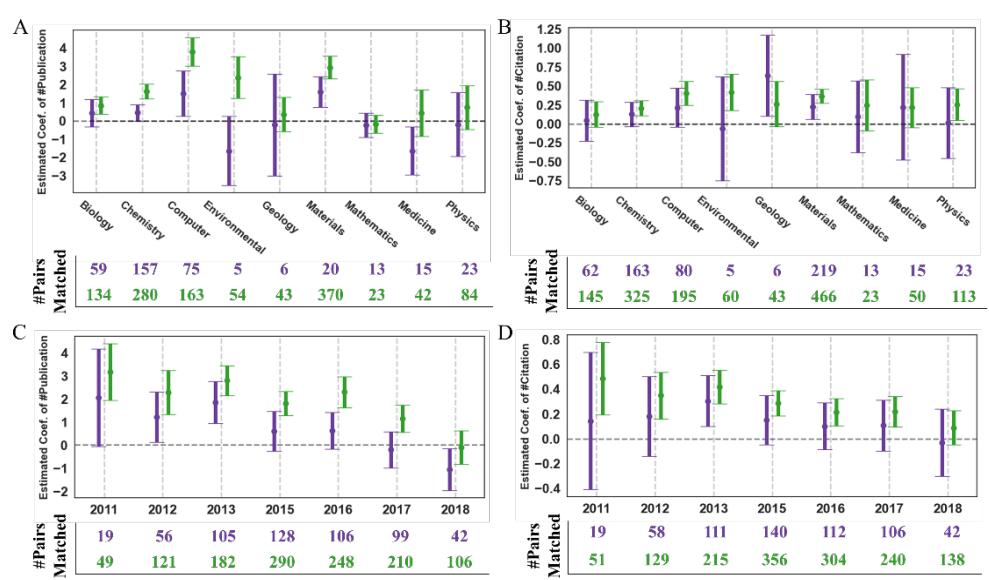

**Figure 3. Estimated difference of publications and citations for the talents and the matched counterparts in different disciplines and movement years.** Panels **A** and **B** show the estimated coefficients of publications and citations in different disciplines, and panels **C** and **D** are the results in different movement years. Error bars show the 95% CI. The tables following each panel represent the number of matched pairs in each category.

### 2.3. Potential factors related to researchers' career development.

The main challenge in scientific movement is the dilemma in the changes of scientific environment. The movers will get new affiliations with new colleagues, potential extra fundings, and students, while being faced to the changes of collaboration ties, team reorganization, and even research directions in new fields. To clarify the latent influence, we are trying to quantify how the potential factors' changes related to scientists' future career development. We take the series of key factors which are strongly correlated to scientific performance and can be observed from our data. The factors are the change rates of (1) collaborators $D_A$, (2) range of collaborative institutions $D_I$, (3) topic directions $D_C$ (see Method section for definitions). We also introduced (4) the difference in team size and further controlled the career start year $Y_0$, discipline, the movement year $Y_w$, and the group which scientists belong to (see Method section and section 3.3 in SI for the details). We use logistic regression models to quantify how these factors correlate with the future scientific output of talent group scientists and their matched unmoved and moved (to China) scientists in the first step of the two-step match procedure. The dependent variable is a binary one determined by whether the scientist's future number of publications (or citations) is above (1) or below (0) the median among their peers.

**Table 2. Logistic regression on the development situations of scientists from different groups.** The outcome variables equal to 1 when the scientists' publication (or citation) counts in five years after movement preponderates over the median of the whole talent pool. The estimated coefficients, standard errors (SEs), and 95% confidence intervals (CIs) are reported for each predictor.

|  | Success in Publications | Success in Citations |
|---|---|---|
| Rate of collaborators change ($D_A$) | 3.4458*** (0.0256) | 1.2286*** (0.0208) |
| Rate of collaborative institutions change ($D_I$) | 2.5878*** (0.0171) | 1.6090*** (0.0154) |
| Rate of research direction change ($D_C$) | -0.2456*** (0.0202) | -1.6338*** (0.0183) |
| Team size change ($\Delta$Team Size) | -0.0570*** (0.0010) | -0.0048*** (0.0008) |
| Career start year ($Y_0$) | Yes | Yes |
| Year of movement ($Y_w$) | Yes | Yes |
| Discipline | Yes | Yes |
| Group | Yes | Yes |
| #Obsevations | 440,350 | 440,350 |
| $R^2$ | 0.2143 | 0.0996 |

*p<0.05, **p<0.01, ***p<0001

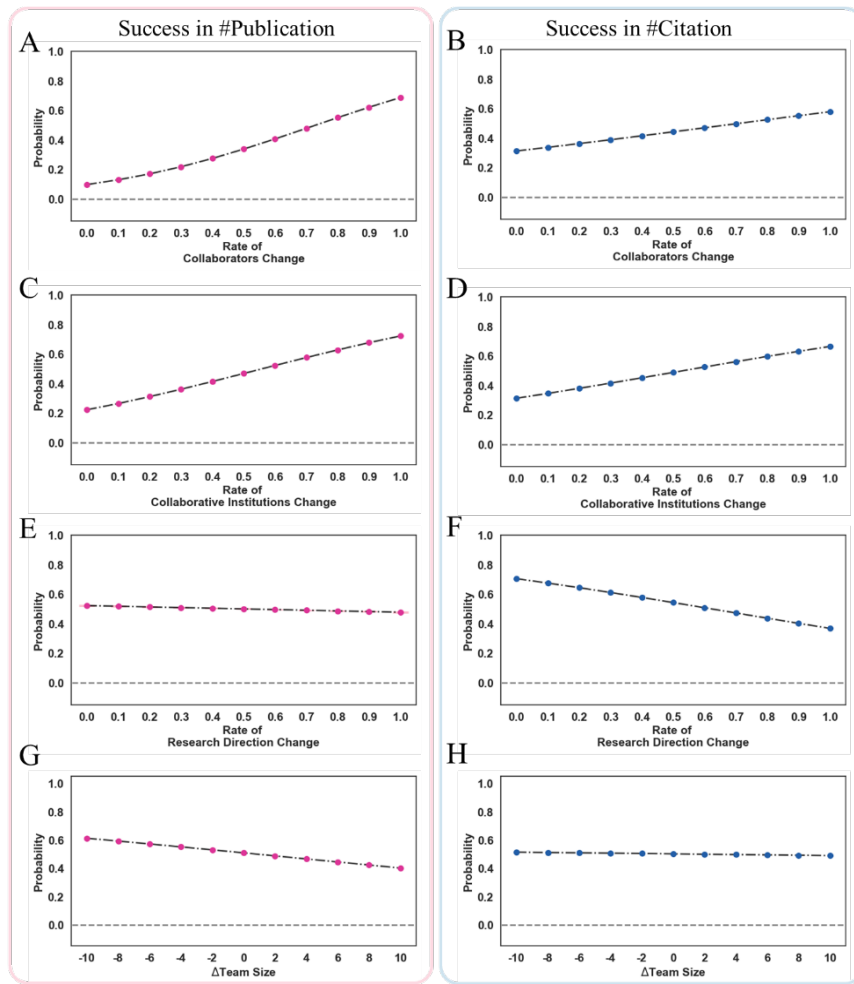

**Figure 4. Model estimation of future scientific performance.** The x-axes show the change rate of different scientific factors: collaboration (**A** and **B**), institution (**C** and **D**), research direction (**E** and **F**) and team-size (**G** and **H**). The y-axes show the success probability of the scientist's future performance is above the median performance. Dot-dashed lines show the estimated trends, and the error bars show the 95% confidence intervals.

Table 2 shows that expanding the collaboration network by adding new collaborators and institutions increases the probability of future scientific success. However, changing the research direction has a negative effect on both publications and citations (the estimated coefficient for $D_C$ is -0.2456 with p-value < 0.001 and -1.6338 with p-value < 0.001 respectively). This is because the talented scientists who switch to new fields lose their leadership and reputation advantages in their original domains. To gain recognition in new areas, they need time to be spotted. In Fig. 4, we further investigate how the probability of future success (publication or citation in the top 50%) varies with different scientific factors. We find that if more than 80% of a talent's collaborators are different from those before mobility, they have a 50% or above chance of success in publications and citations (Fig. 4A and 4B). The change of research direction leads to a slight decrease in productivity and a large decrease in citations, and the increase of the average number of coauthors per paper (team size) reduces the success rate in publications.

## 3. Discussion

We investigated the scientific performance of cross-border young talents who moved to China and received generous support from the government and their host institutions. Using high-resolution data on publications and citations, we compared them with two control groups: scientists who moved to China without the talent hat and scientists who did not conduct cross-border movements. We show that the funded talents outperformed their peers in both productivity and citation impact of research output after moving to China. In this regard, our research has provided compelling evidence of the significance of cross-border talent mobility and support programs in the era of globalization. This not only contributes to the internationalization of China's research ecosystem but also serves as a successful model for scientific collaboration and knowledge exchange on a global scale.

Our results highlight the role of start-up funding for early career scientists, especially in experimental disciplines that require more resources. However, since the Young Thousand Talent program only covered a small fraction of young scientists (about 4,000), we suggest that more support should be given to other young scientists who did not receive this kind of funding. This would help to foster a more inclusive and diverse scientific community in China. We also show that the talent program has less impact for some theoretical disciplines, implying the need for discipline-specific career development policies. However, it is important to acknowledge that our analysis is constrained by the absence of comprehensive data on other forms of research funding. We face constraints in quantifying resources, future research endeavors that encompass a broader range of funding sources would enable a more comprehensive understanding of the intricate interplay between funding mechanisms and scientific productivity among early career scientists in different disciplines.

Our work raises an open question for cultivating cross-border scientists and science development. For the younger PIs who are just independent from their collaborators and mentors, inevitably, they need to change their research directions, especially for cross-border scientists whose research environments undergo significant shifts when they relocate. Such transitions align with the general pattern observed in science, often referred to as the "pivot penalty," where researchers may face challenges and productivity dips when they switch their research focus. Despite the initial challenges, these shifts offer opportunities for growth, interdisciplinary collaboration, and innovative contributions to science([Hill, Yin et al. 2021](#)). Exploring new research directions is necessary for disruptive science and should be encouraged, however, we found that changing research directions after moving was associated with a decrease in both publications and citations, suggesting a risk for young scientists who are newly independent from their collaborators and mentors. We recommend that the host institutions provide more guidance and flexibility for young talents to establish their research agendas in a new environment, for instance, extending the tenure track probationary period.

Moreover, we observed that young scientists who downsized their teams after moving to a new institution were more likely to achieve higher levels of productivity (Fig. 4G and 4H reflect this trend), with smaller team sizes after the cross-border move being associated with increased productivity. Previous studies showed smaller teams tend to introduce more disruptive ideas (Wu, Wang et al. 2019). This implies that downsizing teams may allow young scientists to challenge established paradigms and achieve greater success in their research careers, which, in return, can also serve as an internal motivator for increasing their productivity.

Although the rigorous of our Synthetic control methods (SCM) to evaluate the effects of mobility and funding policies on the academic performance of talented scientists, it still has limitations. SCM focus on quantifying the policy effects but has limited explanatory power for the in-depth causal paths. With supplement data sources, future attempts would focus on combining SCM with other analytical methods or qualitative designs to provide a more comprehensive explanatory framework, especially for policy decision-making. Another potential challenge in this study, and in many others of its kind, is the unobservable factors. This quantitative study can control measures for observable variables, such as publication numbers and citations, but may not capture researchers' invisible ability adequately. However, the unobservable and observable variables are usually correlated, we control the observable ones could partially controlled the unobservable one, future studies would focus on integrating high-resolution data to further improve the control and the assessments.

## 4. Method

### 4.1. The Synthetic Control Method (SCM)

The SCM was used here to eliminate potential interference that could have been induced by individual qualities such as field, affiliation, reputation, mentorship, etc. (Abadie and Gardeazabal 2003, Abadie, Diamond et al. 2015). The counterfactual in SCM is a weighted average of the potential control groups. Assume the counterfactual of one aspect of the talent's academic performance is:

$$Y_{i,t}^{CF} = \sum_{j=1}^{N_i} weight_j \times Y_{j,t}. \qquad (1)$$

$N_i$ is the set of candidates benchmarked with winner $i$, $j$ is one of $N_i$, and $Y_{j,t}$ is the academic performance of candidate $j$. Then the pre-mobility difference between the talent group and the counterfactual is:

$$\Delta_{Y,pre} = Y_{pre}^T - Y_{pre}^{CF} = Y_{pre}^T - \sum_{j=1}^{N_i} weight_j \times Y_{j,pre}. \qquad (2)$$

The counterfactual is obtained then by minimizing the function of $\Delta_{Y,pre}$. To synthesize control units in scale of publication counts, the yearly publication counts for the past 5 years and the mean citation count in the 5 years before mobility were considered. The contender groups $G_1$ and $G_2$ were generated for the talent group $G_w$ based on the yearly increased publications and citations, respectively. The widely-used tool for SCM in Stata was used to synthesize control units for each treated scientist in the variables of interest (Abadie, Diamond et al. 2011).

### 4.2. Difference-in-Difference (DID) regression

To evaluate the effects of talent hat and mobility on academic performance, we used a DID regression model, which was modified to include fixed time and individual effects. The model was used to estimate the extent to which a talent scientist's scientific production and citation exceeded their peers after movements. We denote time relative to the movement year as $t$ ($t = 0$ is the movement year) and define $Post_t$ as the dummy variable of the year period. The mathematical formula of this model is:

$$Y_{st} = \beta_0 + \beta_1 \cdot Treat_s \times Post_t + \mu_s + \tau_t + \epsilon_{st}. \quad (3)$$

Where $\beta_1$ is the coefficient for the cross term, $Treat_s \times Post_t$, which is a dummy variable equaling 1 for observations of talent' academic measurements in the posterior mobility period (otherwise it is zero) and $\epsilon_{st}$ is the error term. $\mu_s$ denotes individual fixed effect for scientist $s$ while $\tau_t$ denotes time fixed effect in scale of year (See section 3.1 in SI).

### 4.3. Predict the success of talent scientists

In the logistic regression of the development situations of scientists, we introduce four measures, $D_A$, $D_I$, $D_C$ and $D_{size}$ to represent collaborators' dissimilarity, collaborative institutions range, research direction (each paper's research direction is identified by the topic levels defined in OpenAlex), and change of team size (number of coauthors within a paper). Mathematically, we denote the collaborators of talent before and after movement as set $A_0$ and set $A_1$ respectively. Then $D_A$ is calculated by:

$$D_A = \frac{|A_1 - A_0|}{|A_1|} \quad (4)$$

Where $|\cdot|$ is the mode of a set and '−' is the set subtraction. The calculation is similar for $D_I$ and $D_C$. $D_{size}$ is the difference between the average of team size for a scientist after and before movement, which is: $D_{size} = Size_1 - Size_0$. The model formula is as the following:

$$\ln\left(\frac{Prob_i}{1-Prob_i}\right) = \beta_0 + \beta_A D_A + \beta_I D_I + \beta_C D_C + \beta_S D_{size} + \lambda_i Displine_i + \mu_w Y_w + \mu_0 Y_0 + \gamma_i Group_i + \epsilon_i.$$

(5)

$Prob_i$ is the probability that scientist $i$'s publication or citation counts in the following 5 years after movement exceed the median of the entirety of $G_w$. $Y_w$ is the year of movement. And $Y_0$ is the year when scientist $i$ published the first paper which means the start of one's career. $Displine$ is her/his research direction of interest. $Group$ is the group which she/he belongs to. (Find more details in section 3.3 of SI).

**Acknowledgments:** This work was supported by the National Natural Science Foundation of China (grants No. NSFC62006109 and NSFC12031005), and partially supported by the Support Plan Program of Shenzhen Natural Science Fund under Grant 20220814165010001. The computation in this study was supported by the Center for Computational Science and Engineering of SUSTech.

**Author Contributions:**
Y. Huang: Data curation, Visualization, Investigation, Writing—Original draft preparation.
X. Cheng: Formal analysis, Writing—review& editing.
Ch. Tian: Validation, Writing—review & editing.
X. Jiang: Data curation, Validation, Writing—review & editing.
L. Ma: Data curation, Validation, Writing—review & editing.
Y. Ma: Conceptualization, Supervision, Project administration, Writing—review &editing.

**Competing Interests:** The authors declare no competing interests in this work.

**Data and Code Availability:** The OpenAlex data is publicly available. The code for reproducing the main results in this work are available in the supplementary materials.

# Supplementary Information for

# Talent hat, cross-border mobility, and career development in China

Yurui Huang[1], Xuesen Cheng[2], Chaolin Tian[1], Xunyi Jiang[1], Langtian Ma[1], Yifang Ma[1*]

[1]Department of Statistics and Data Science, Southern University of Science and Technology, Shenzhen 518055, Guangdong, China

[2]Center for Higher Education Research, Southern University of Science and Technology, Shenzhen 518055, Guangdong, China

*Email: mayf@sustech.edu.cn

# CONTENTS

## Materials and Methods

## Supplementary Figures and Tables

# 1. Data processing

## 1.1. Scientific corpus data

In this work, we used the large-scale publicly available scientific corpus, OpenAlex (https://openalex.org). The OpenAlex database provides integrated multidimensional scientific data including the Microsoft Academic Graph, Crossref, Pubmed, etc. (Priem, Piwowar, and Orr 2022) and contains detailed information on various entities such as works, authors, institutions, and journals, each identified with unique identification numbers. Additionally, it includes information on links between those entities including authorship links, collaboration, citing-cited relationships, and host affiliations, which enables us to conduct a thorough quantitative analysis. OpenAlex does a sophisticated name disambiguation process for authors using multiple features from the data including coauthor's information (name, affiliation, etc.), citation network, and other features of the data. Due to its comprehensive and available information on scientific data, OpenAlex dataset has been employed to investigate a variety of questions in the field of the science of science (Nishikawa-Pacher, Heck, and Schoch 2022; Saier, Krause, and Färber 2023; Hao et al. 2022).

After conducting data cleaning and organization on the data extracted from this dataset, we obtained crucial information on the works including their publication dates, author ships, institutions, disciplines. We then integrated this data on works by authors to compute the number of articles published and new citation counts for each author per year. To determine the primary research interests of each scientist, we concentrated on the most frequently occurring primary disciplines among all the articles authored by each scientist. Finally, we focused on the geographical attributes of the institutions where the scientists have publication records to gain insights into their work locations and timelines.

The OpenAlex database includes data from nearly 60 million authors who have published at least one article. In our study, we evaluate the synergistic effects of talent program and migration on scientists' career development by adapting observational studies method (Altmann 1974; Jepsen et al. 2004), based on whether they are being elected into one of China's oversea younger talent program. As the scientists in the talent program are relatively young scholars (as one of the requirements for applicants for this plan is to be less than 38 years old), we ensured that all scientists included in the talent group and potential contender groups had published their first article no earlier than 2000 before the matching process. Additionally, we only considered scholars who have published a minimum of 10 articles to increase the accuracy of the matching algorithm applied in this work and reduce the estimation errors of the regression models.

## 1.2. Cross-border talent dataset

The Young Thousand Talents Program is a Chinese government-sponsored initiative aimed at attracting and nurturing top young talent across various fields, with a focus on advancing scientific and technological development in China (Cao et al. 2020; Jia 2018; Lewis 2023; Lundh 2011; Shi, Liu, and Wang 2023). The program, which was launched in 2011, primarily targets young Chinese scholars and researchers who have achieved significant academic and research accomplishments in their respective fields abroad. Participants in the program are eligible for a range of benefits, including higher pay, generous research funding, and other forms of support to help establish their research careers in China.

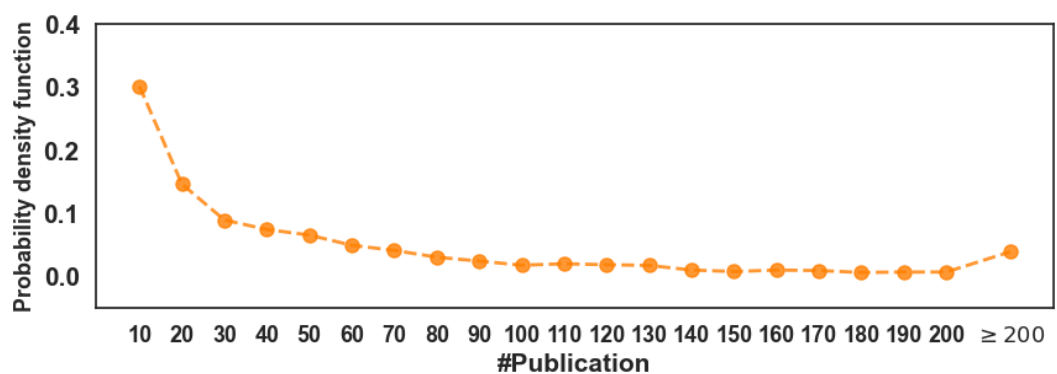

**Figure S1.** Distribution of number of publications in the talent group.

We manually curated the list of talents in from 2011 to 2018 from the announcement websites (2014 year are missing) which includes the scholar's name, affiliation, year, etc. Then for each talent, we manually collect her/his publications by searching publicly available information, for most of the cases, we can locate the scholar's current university homepage which has publication records, ORCID link, google scholar link etc. Finally, we supplemented the scholar's publication records using the OpenAlex.

As a result, we successfully matched a total of 2589 scientists, with the majority having published less than 100 articles throughout their careers (as illustrated in Figure S1). To minimize systematic errors, we included only those scientists whose first publication was between the years 2000 and 2015 and have at least 10 publications. Our final analysis included 1563 STEM scientists as talent ones who were elected to the talent program between 2011 and 2018, primarily in the fields of computer science, materials science, biology, chemistry, physics, mathematics, environmental science, geography, and pharmacy (medical science) (as demonstrated in Figure S2 A and B).

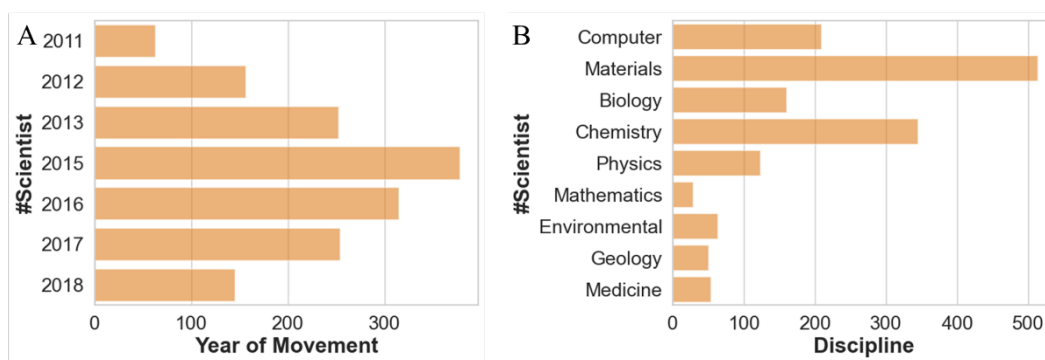

**Figure S2.** The number of talents studied by years of movement and by disciplines. (A) years of movement; (B)disciplines

## 1.3. Matching contender groups

We employed two distinct contender groups of scientists in our study to analyze the cross-border movement effects and the talent effects. The first group, $G_1$, comprises scientists who experienced displacement but did not receive support from the talent program. Displacement is defined as the occurrence when a scientist, who has been publishing their research in non-Chinese institutions for a continuous period of three years or more, begins publishing in Chinese institutions ([Smith 2009](); [Van Noorden 2012](); [Edler, Fier, and Grimpe 2011]()). The displacement time is recorded as the year of the first publication in a Chinese institution. Before matching, there were 103,698 scientists in $G_1$, with movement years ranging from 2000 to 2021 (see Figure S3A). In Figure S3B, we showed the number of scientists in the top 10 countries (ranked by numbers) who move to China afterwards.

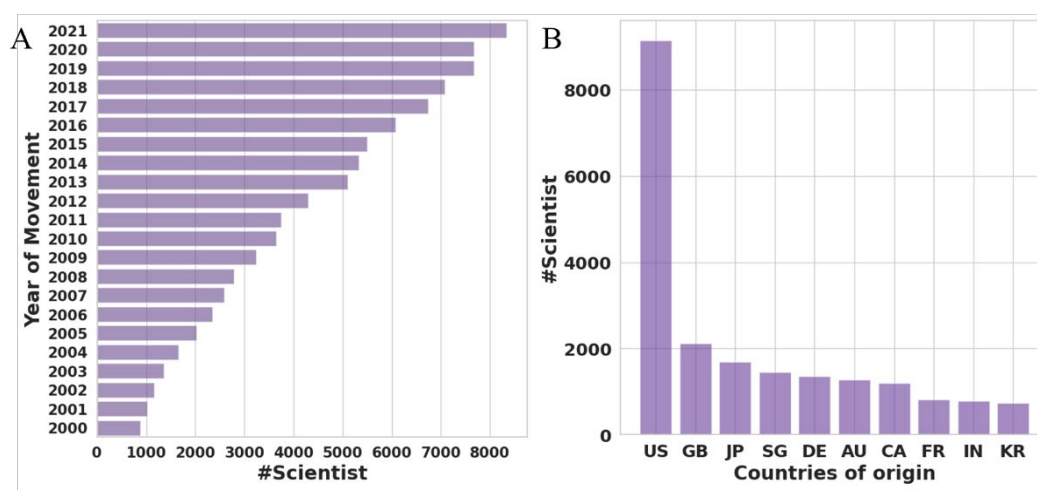

**Figure S3.** The number of scientists who moved to China without enrollment in the talent program studied by years of movement and by their original studying countries. (A)years of movement; (B) original studying countries.

The second group, $G_2$, consisting of scientists who did not move, i.e., they published their papers with affiliations in the same country. The original $G_2$ group consisted of over 10,000,000 scientists publishing their first papers in time ranging from 2000 to 2021 and in different countries (see Figure S4).

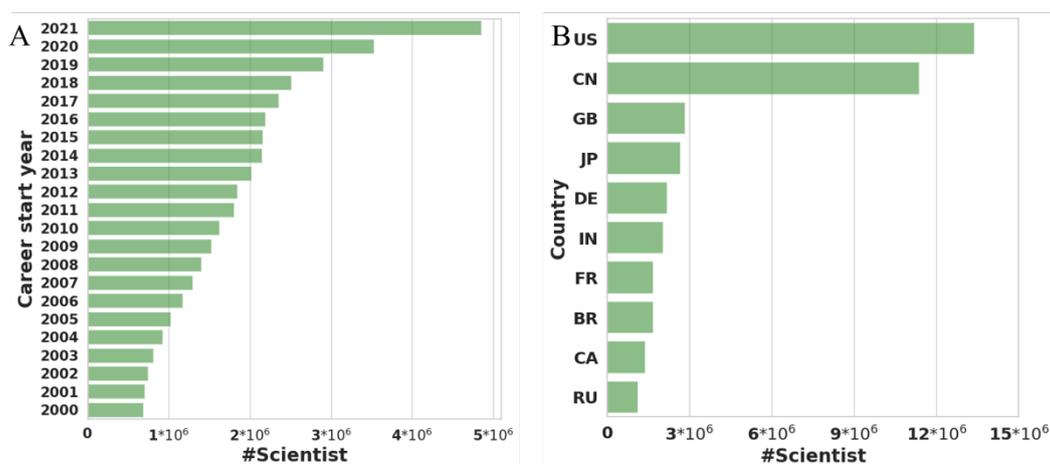

**Figure S4.** The number of scientists who did not conduct cross-border movements by the career start year and by the affiliated countries. (A) career start years; (B) affiliated countries.

## 2. Matching Procedure

### 2.1. Two-step matching for contender groups

**"Exact" matching step.** At first, the "exact" step, we conducted a match, i.e., for each cross-border talent scientist, we matched 300 unmoved scientists and 200 moved (to China) scientists without enrollment in talent program who have the same discipline, close research career starting year, and comparable total number of publications and citations in the five years prior to movement with the talent scientist. Specifical for unmoved scientists, we request that the year of their last publication must exceed that of movement of their matched talent scientists. For moved scientists then we request that the year of their movement is close to their matched talents. These measures are intended to ensure the comparability between talent group and its contenders.

**"Refining" step.** Secondly, the "refining" step, we further matched the identical scientists by considering the yearly publications and yearly citations to make sure they display the same scientific development curve prior to the talent recruitment year. In this step, we leveraged the established matching techniques. We used the synthetic control method (SCM) in the main results. The SCM is a statistical technique that constructs a "synthetic" control group, which closely matches the characteristics of the treated unit, except for the fact that it did not receive the treatment (Abadie, Diamond, and Hainmueller 2015). This approach is used to solve the problem of matching similar control units to the treated ones in observational research(Abadie, Diamond, and Hainmueller 2010; McClelland and Gault 2017). The method involves constructing a weighted average of a set of control units that serve as the counterfactual for the treated unit. The weights are chosen in such a way as to minimize the distance between the pre-treatment outcomes of the treated unit and the synthetic control group, subject to some restrictions on the weights. The SCM produces 'virtual' scientists for each treated scientist, rather than matching them with real scientists. This allows us to estimate the treatment effect accurately and address the issue of selection bias.

Assume the counterfactual of one aspect of $G_w$'s academic performance is:

$$Y_{i,t}^{CF} = \sum_{j=1}^{N_i} weight_j \times Y_{j,t} \qquad (1)$$

$N_i$ is the set of candidates benchmarked with winner $i$, $j$ is one of $N_i$, and $Y_{j,t}$ is the academic performance of candidate $j$. Then the pre-treatment difference between the treatment group and the counterfactual is:

$$\Delta_{Y,pre} = Y_{pre}^T - Y_{pre}^{CF} = Y_{pre}^T - \sum_{j=1}^{N_i} weight_j \times Y_{j,pre} \qquad (2)$$

The counterfactual is obtained then by minimize the function of $\Delta_{Y,pre}$.

We applied this two-step matching procedure to identify indistinguishable control groups from the millions of scientists in the database. In order to assess the effectiveness of this procedure, we separately evaluated the differences of academic performance in the pre-mobility period between the talent group and contender groups at different steps. As shown in Figure S5, we could tell that the "refining" contender group ($G_1$ and $G_2$) did not differ significantly from the talent group in terms of the number of publications and citation before mobility. In contrast, the "exact" contender group ($G_1^*$ and $G_2^*$) significantly differed from the talent group ($G_w$) at almost all time points, making it impossible to unbiasedly estimate the effect of mobility and talent program through difference-in-difference regression. Therefore, the SCM is a useful tool for estimating the effects of treatments or interventions in situations where randomized experiments are not feasible.

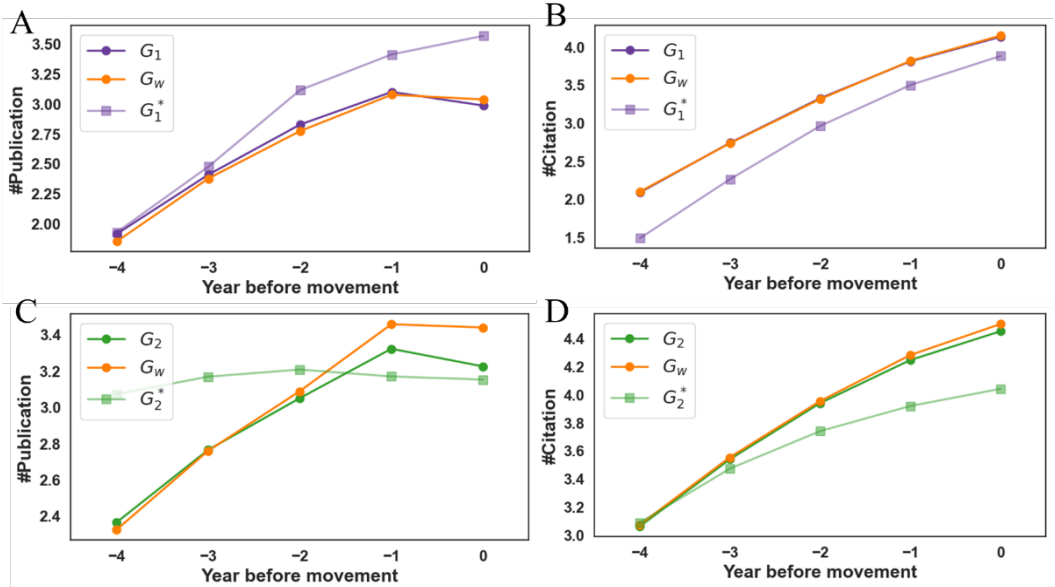

**Figure S5.** Comparison on academic performance measurements in the pre-mobility period of contender groups at different steps of the two-step matching. The 'refining' step improve the effectiveness of matching so that bias casued by the prior difference

between contender group and talent group could be eliminated. $G_1^*$ and $G_2^*$ are contender groups at 'exact' step while $G_1$ and $G_2$ are contender groups at 'refining' step.

**Table S1.** Present academic measurements in the pre-mobility period of talent scientists who moved to China compared with contender groups who moved without talent hat at difference steps of the two-step matching in a balance table.

|  | Publications | | Citations | |
| --- | --- | --- | --- | --- |
| Year | 'exact' | 'refining' | 'exact' | 'refining' |
| -4 | -0.0763 (0.0980) | -0.0626 (0.1121) | 0.6112*** (0.0600) | 0.0114 (0.0767) |
| -3 | -0.0988 (0.1129) | -0.0318 (0.1369) | 0.4716*** (0.0629) | -0.0047 (0.0715) |
| -2 | -0.3424** (0.1259) | -0.0541 (0.1439) | 0.3522*** (0.0583) | -0.0110 (0.0659) |
| -1 | -3650** (0.1414) | -0.0514 (0.1630) | 0.3167*** (0.0515) | 0.0062 (0.0598) |
| 0 | -0.6245*** (0.1563) | -0.0419 (0.1848) | 0.2667*** (0.0462) | 0.0159 (0.0573) |
| #Observations | 411778 | 5550 | 411995 | 5880 |

$^*p<0.05, ^{**}p<0.01, ^{***}p<0001$

**Table S2.** Present academic measurements in the pre-mobility period of talent scientists who moved to China compared with contender groups who did not conduct cross-border mobility at difference steps of the two-step matching in a balance table.

|  | Publications | | Citations | |
| --- | --- | --- | --- | --- |
| Year | 'exact' | 'refining' | 'exact' | 'refining' |
| -4 | -0.7593*** (0.0922) | -0.0520 (0.0946) | -0.0153 (0.0337) | 0.0114 (0.0533) |
| -3 | -0.4295*** (0.0957) | -0.0257 (0.1089) | 0.0783* (0.0307) | 0.0116 (0.0482) |
| -2 | -0.1731** (0.0992) | -0.0153 (0.1191) | 0.2111*** (0.0295) | 0.0116 (0.0455) |
| -1 | 0.1619 (0.1030) | 0.0094 (0.1364) | 0.3597*** (0.0292) | 0.0326 (0.0440) |
| 0 | 0.0624* (0.1078) | -0.0104 (0.1588) | 0.4584*** (0.0292) | 0.0483 (0.0441) |
| #Observations | 2064314 | 12080 | 2066383 | 14330 |

$^*p<0.05, ^{**}p<0.01, ^{***}p<0001$

## 2.2. Supplementary matching methods

The main method used in this study is the SCM, which is employed to identify an appropriate control group for the treatment group in an observational study. Unlike traditional methods, we synthesized a virtual group of scientists using a linear transformation instead of finding a group of them that are most similar to the existing talent group in the data. The successful matched pairs are small for the $G_w$ and $G_1$ groups because of the lack of quantity of scientists who moved to China without enrollment in talent program. To demonstrate the robustness of our conclusions, we introduce two additional methods: the Coarsened Exact Matching (CEM) ([Blackwell et al. 2009](); [Iacus, King, and Porro 2012]()), and a newer matching method called Dynamic Optimal Matching (DOM) inherited from Optimal Matching method ([Pimentel et al. 2015](); [Rosenbaum 1989]()). These two matching methods have been used in recent scientific research ([Jin, Ma, and Uzzi 2021](); [Bertoni et al. 2020](); [Guarcello et al. 2017](); [Stevens, King, and Shibuya 2010]()) and are compared with the SCM method in the performance of finding comparable contender groups in our work.

CEM is a popular statistical method in practical research due to its well-developed packages in several programming languages such as Python, R, and Stata. The CEM algorithm creates a set of strata, and units in the strata that contain at least one treatment, and one control unit are preserved in the sample, while units in other strata are deleted. Strata can be created and chosen using a variety of criteria and algorithms. In this research, the counts of publication and citation in each prior timestamp are the covariates in the formula of the CEM model. In this case, the between-group variance is the difference between the means of the two groups of participants.

The DOM technique consists of two steps. The first step involves selecting a matched candidate pool. Specifically, for each $G_w$ scientist $i$, we selected up to 40 close-distance candidate scientists based on a distance measure $\theta_{i,j}$ from the $G_1$ group of scientists who moved to China without support from talent program:

$$\theta_{i,j} = \frac{\sum_{t=-4}^{0}\left[\left(\log P_{i,t} - \log P_{j,t}\right)^2 + \left(\log C_{i,t} - \log C_{j,t}\right)^2\right]}{12} \quad (3)$$

where $P_{i,t}$ and $C_{i,t}$ are numbers of publication and citation of scientist $i$ at the year before movement $t$. Following this, we used the matching covariates, including the number of publications and citations within the period of four to zero year prior to movement, as inputs in the optimal matching using the MatchIt package in R to identify a matched contender scientist who moved to China without enrollment in talent program for each talent scientist (if possible) ([Stuart et al. 2011]()). We then estimated the average treatment effects in the treated group (ATET) for talent group compared with contender groups found by both CEM and DOM using DID regressions.

The CEM and DOM method offer us more (almost double) matched pairs of talent scientists and contender who moved to China without talent program than SCM method. By conducting DID regression on academic performance of them, we found much the same results derived from the comparison mentioned in the main context, i.e., the talents' numbers of publications and citations exceed the group without talent hats even if both groups performed displacement to China.

Table S3. Estimated coefficients of DID regression on talent group versus contender groups selected by CEM and DOM. The regression coefficients are displayed with their corresponding significance levels, followed by the standard errors in parentheses.

|  | CEM | | DOM | |
| --- | --- | --- | --- | --- |
| Measures | Publications | Citations | Publications | Citations |
| Talent hat×Movement | 0.4270** (0.1306) | 0.2557*** (0.0249) | 0.3137* (0.1579) | 0.2537*** (0.0350) |
| Individual | Yes | Yes | Yes | Yes |
| Time | Yes | Yes | Yes | Yes |
| #Pairs Matched | 1082 | 1083 | 1155 | 1155 |
| N | 182538 | 182538 | 59904 | 21597 |
| $R^2$ | 0.4720 | 0.8359 | 0.5479 | 0.8539 |

*p<0.05, **p<0.01, ***p<0001

# 3. Variables and Model Specification

## 3.1. Difference-in-Difference regression

The difference in differences (DID) regression model is a statistical method used to estimate the treatment effect by comparing changes in outcomes over time between treatment group and control group. This method has found wide application in various fields, including economics, health sciences, and social sciences, and has recently been adapted for use in the science of science(Branas et al. 2011; Conley and Taber 2011; Roth et al. 2022; Huang, Tian, and Ma 2023).

Mathematically, the DID model estimates the average treatment effect by taking the difference in the average outcome between the treated and control groups before and after the treatment, and then subtracting the difference in the average outcome between the two groups in the pre-treatment period. The practical application of the DID model involves several steps. First, researchers must identify a treatment group and a control group that are similar in terms of observable characteristics that might affect the outcome of interest. Second, they must collect data on the outcome variable for both groups in the pre-treatment and post-treatment periods. Third, researchers must estimate the DID model using the collected data to obtain the treatment effect estimate.

Denotes the time relative to the year of movement as $t$, $t$ is defined ranging from -4 to 9. $Post_t$ is the dummy variable indicating the moved/unmoved period, where $Post_t = 0$ indicates the time interval that $t$ less than or equal to 0 and $Post_t = 1$ for $t$ ranging from 1 to 9. Use the number of publication and the number of citations as measurements on scientists' academic performance. We conducted DID regression with fixed time effect and individual effect. The goal of time effect control is to reduce the influence of unobservable homogenous shocks in the temporal dimension. Meanwhile, we used the fixed individual effect in case there were missing factors that affected scientist's academic performance. The mathematical formular of this model is:

$$Y_{st} = \beta_0 + \beta_1 \cdot Treat_s \times Post_t + \mu_s + \tau_t + \epsilon_{st}, \quad (4)$$

Where $\beta_1$ is the coefficient for the cross term, $Treat_s \times Post_t$, which is a dummy variable equaling 1 for observations of talent's academic measurements in the post-movement period (otherwise it is zero) and $\epsilon_{st}$ is the error term. $\mu_s$ denotes individual fixed effect for scientist $s$ while $\tau_t$ denotes time fixed effect in scale of year.

## 3.2. Yearly effects of movement and talent hat

To estimate the yearly effect, we introduced variables in our regression model that allows us to control for any underlying trends in the outcome variable over time. We can then estimate the effect of movement and talent hat separately for each year after mobility to see how it varies over time. Specifically, we operationalize the $Post$ variable by specifying it as the number of years after mobility, i.e., $Post^1$, $Post^2$, ..., $Post^9$. Therefore, Formula (4) is modified accordingly:

$$Y_{st} = \beta_0 + \mu_s + \tau_t + \epsilon_{st} + \sum_{T=1}^{9} \beta_{1T} \cdot Treat_s \times Post_t^T, \quad (5)$$

$Treat_s \times Post_t^T$ are a series of dummy variables equaling to 1 for observations of talent's academic measurements in the $T$th year after mobility (otherwise it is zero). Hereby $\beta_{1T}$ are the coefficients for the yearly cross term, of which estimations represents the yearly effect of movement and talent hat.

Career development of scientists can be influenced by various factors, especially their research of interests. In this context, the impact of mobility and talent program on career development can be examined among scientists from different disciplines and the result is shown in Figure S6. Movement can have a positive impact on the career development of scientists, as it allows them to gain new experiences, collaborations and expand their professional network. However, it varies across disciplines, as the nature of the research and academic culture may differ. For instance, mobility may be more effective in interdisciplinary fields, where collaborations across different disciplines are more common. Moreover, talent program can also have a positive impact on the career development of scientists, as it can provide recognition for their research and increase their visibility in the academic community. This can lead to more funding opportunities, collaborations and invitations to conferences and seminars.

As shown in Figure S7, we found that mobility had a detrimental impact on the number of articles published initially. However, for scientists who chose to migrate, although their productivity in terms of the number of publications was affected, they experienced a surge in academic influence power, i.e., citation impact, after migration. The initial negative effect of mobility on the number of articles published suggests that scientists who relocated would encounter with challenges in adjusting to new surroundings and building new collaborations. Nevertheless, the fact that these migrating scientists had an increase in citation after migration indicates that the movement may have led to new opportunities and collaborations that amplified their academic impact. These findings may be used to develop policies related to scientific mobility and recognize potential benefits and challenges related to it that are in line with the findings obtained in the main text.

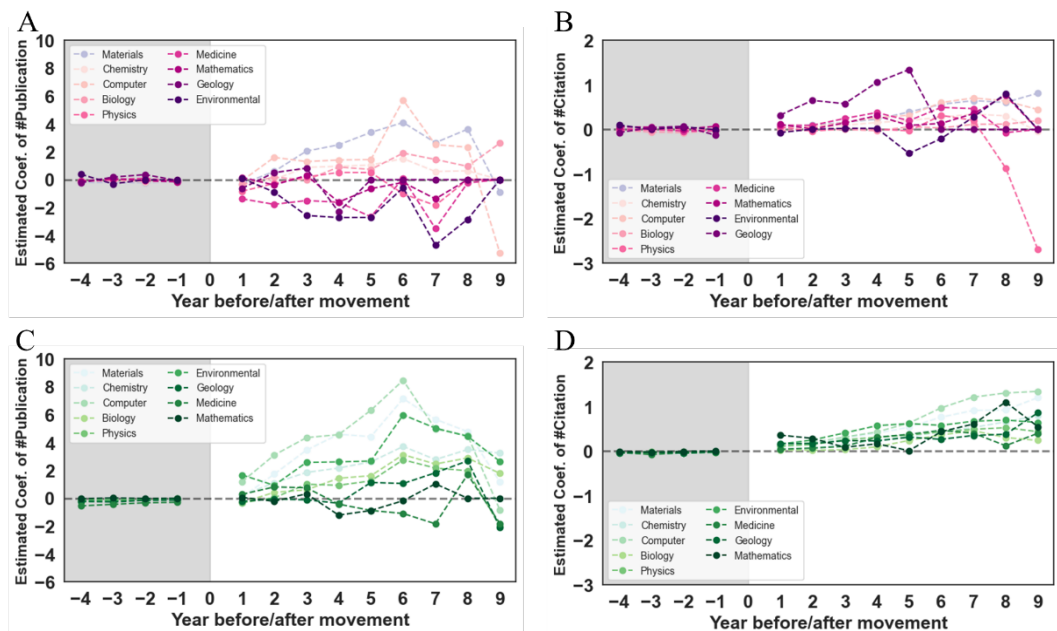

**Figure S6.** Coefficients of DID regression versus different fields. A and C are estimated coefficients of yearly corss term for trials $G_w$ vs. $G_1$ and $G_w$ vs. $G_2$ respectively when the measurement is number of publications, While B and D are estimated coefficients when the measurement is number of citations.

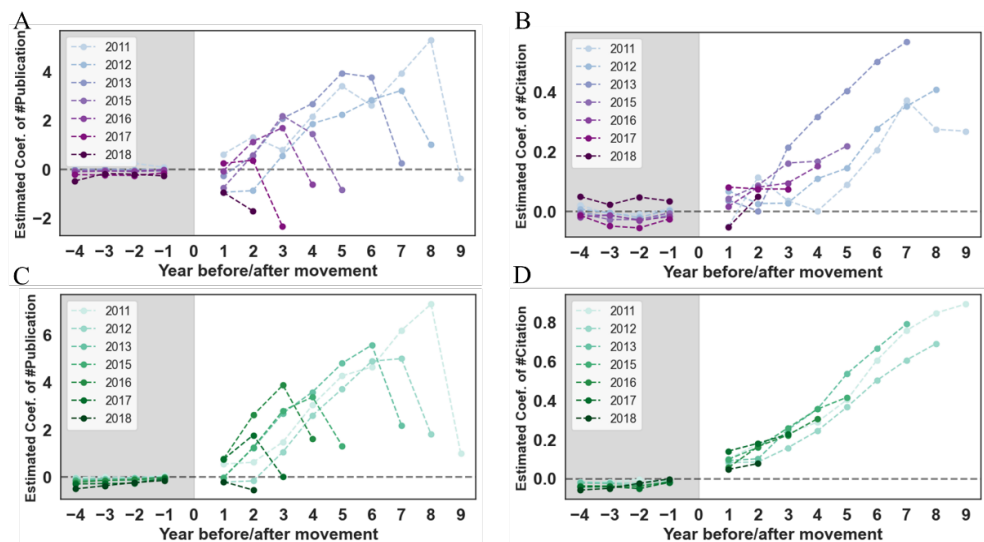

**Figure S7.** Coefficients of DID regression versus different years of movement. A and C are estimated coefficients of yearly corss term for trials G$_w$ vs. G$_1$ and G$_w$ vs. G$_2$ respectively when the measurement is number of publications, While B and D are estimated coefficients when the measurement is number of citations.

### 3.3. Logistic regression of scientists' posterior development situation

Logistic regression is a statistical model used to analyze the relationship between a binary dependent variable and one or more independent variables(Menard 2002). The dependent variable takes on only two possible values, typically coded as 1 (for success) and 0 (for failure). The model estimates the probability that the dependent variable is equal to 1, based on the values of covariates.

In the logistic regression of the development situations of scientists, we introduce four continuous measures as predictors of success probability of G$_w$ scientist, including covariates $D_A$, $D_I$, $D_C$ and $D_{size}$ mentioned in the main context. $D_A$, $D_I$ are the rate of change of collaborative authors and institutions. $D_C$ is the rate of range of the research direction of scientist before and after mobility. The research direction of scientist is determined by the topics of their published papers in some period, which must adhere to the definition set by the OpenAlex of being at a level equal to or greater than 2. $D_{size}$ is the "delta" in the number of coauthors per paper, on average, represents the difference between the before and after mobility for a scientist, which is: $D_{size} = Size_1 - Size_0$. The model formula is as the following:

$$\ln\left(\frac{Prob_i}{1 - Prob_i}\right) = \beta_0 + \beta_A D_A + \beta_I D_I + \beta_C D_C + \beta_S D_{size} + \lambda_i Displine_i + \mu_w Y_w + \mu_0 Y_0 + \gamma_i Group_i + \epsilon_i. \quad (6)$$

$Prob_i$ is the probability that scientist $i$'s publication or citation counts in the following 5 years after movement exceed the median of the entirety of G$_w$. $Y_w$ is the year of movement. And $Y_0$ is the year when scientist $i$ published the first paper which means the start of one's career. $Displine$ is his/her research direction of interest. $Group$ is the group which he/she belongs to.

We also examined the relationship between various variables and corresponding scientific performance indicators among peers after logistic regression using marginal distributions. The results are presented in Figures 4 and S8. The results shown in Figure S8 CD suggest that the year of movement slightly postively impact scientists' success in terms of both productivity and citation impact over the following five years. While the results shown in Figure S8 EF suggest that the research direction do not significantly impact scientists' success in the similar terms.

However, Figure S8 A and B indicates that longer career experience before mobility may be more beneficial for scientists in terms of increased productivity and citation impact. These findings suggest that international research experience could be crucial for scientists' future career development after movement. It is worth noting that further research is necessary to fully comprehend the mechanisms underlying this effect and to explore possible factors that may moderate the relationship between career mobility and success. Researchers affiliated with the distinguished talent group $G_w$ are statistically more inclined to reap advantages in terms of citation impact through mobility when compared to their counterparts in groups $G_1$ and $G_2$. However, it should be noted that they may encounter greater challenges in sustaining their research productivity as compared to their contenders (See Fig. S8 GH).

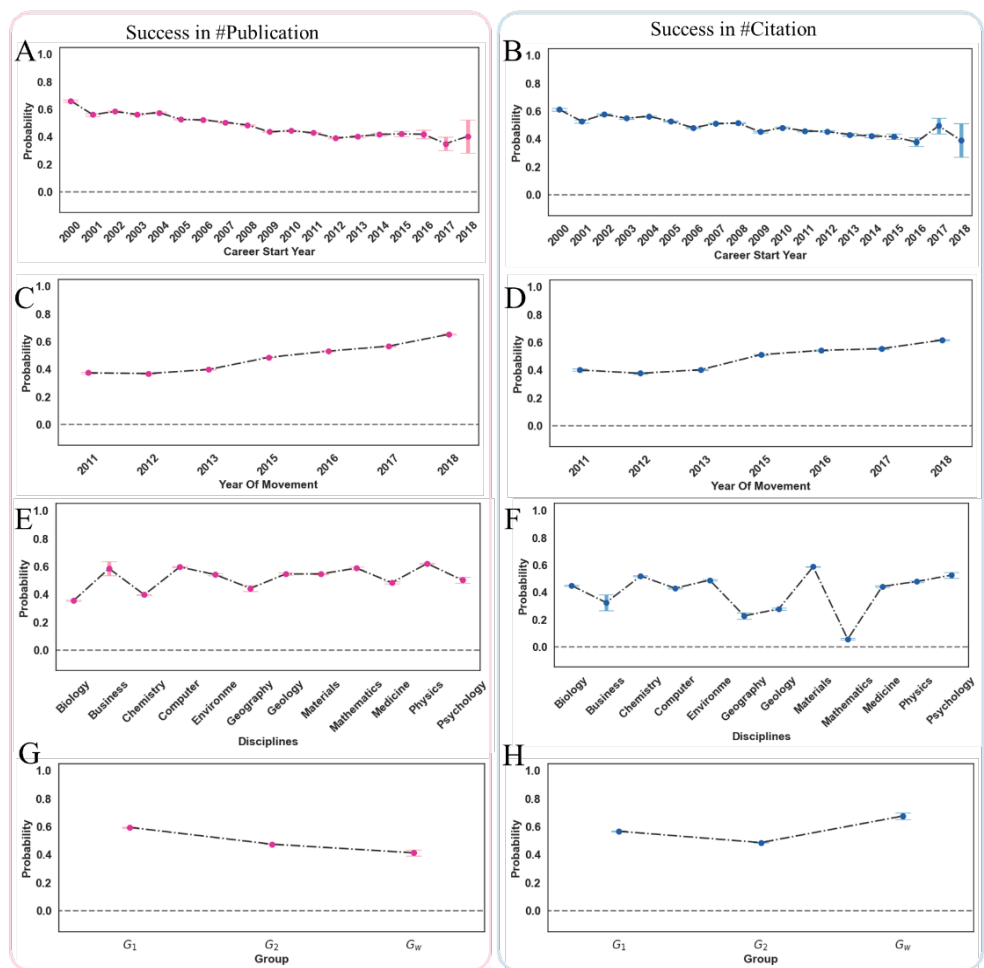

**Figure S8.** Margins plot of logistic regression with discrete covariates. These discrete covariates including Start Years, Return Years and Disciplines. A, C, E and G are estimated coefficients of margins plot for the probability of success in productivity, While B, D, F and H are estimated coefficients of margins plot for the probability of success in citation impact.